\documentclass[12pt]{elsart}
\usepackage{epsfig,graphics}
\usepackage{hyperref}

\newlength{\lslash}

\begin{document}
\begin{frontmatter}
\title{Strangeness in the baryon ground states}
\author[GSI]{A. Semke}
\author[GSI]{and M.F.M. Lutz}
\address[GSI]{GSI Helmholtzzentrum f\"ur Schwerionenforschung GmbH,\\
Planck Str. 1, 64291 Darmstadt, Germany}
\begin{abstract}
We compute the strangeness content of the baryon octet and decuplet states based on an analysis of
recent lattice simulations of the BMW, PACS-CS, LHPC and HSC groups for the pion-mass dependence of the baryon masses.
Our results rely on the relativistic chiral Lagrangian and large-$N_c$ sum rule
estimates of the counter terms relevant for the baryon masses at N$^3$LO. A partial summation is
implied by the use of physical baryon and meson masses in the one-loop contributions to the baryon self energies.
A simultaneous description of the lattice results of the BMW, LHPC, PACS-CS and HSC groups is achieved.
From a global fit we determine the axial coupling constants $F\simeq 0.45$ and $D \simeq 0.80 $ in agreement with
their values extracted from semi-leptonic decays of the baryons.
Moreover, various flavor symmetric limits of baron octet and decuplet masses as obtained by
the QCDSF-UKQCD group are recovered. We predict the pion- and strangeness sigma terms and the pion-mass dependence
of the octet and decuplet ground states at different strange quark masses.
\end{abstract}
\end{frontmatter}

\section{Introduction}

The study of the quark-mass dependence of hadron masses is of great importance to unravel the intricate QCD
dynamics at low energies. It provides a bridge between effective field theory approaches and lattice
simulations. By now, various unquenched three flavor simulations  for the pion-mass dependence of the
baryon ground state masses are available \cite{MILC2004,LHPC2008,PACS-CS2008,HSC2008,BMW2008,Alexandrou:2009qu,Durr:2011mp,WalkerLoud:2011ab}.
Less well studied is the kaon-mass dependence of the hadron masses.

The purpose of this Letter is to present detailed predictions on the strangeness content of the baryon octet and
decuplet  ground states. This is possible using the available lattice data on the pion-mass dependence
of the baryon masses. Given a systematic analysis based on the chiral Lagrangian, the strange-quark mass
dependence of the baryon masses can be calculated. Such results scrutinize the consistency of the chiral extrapolation approach
and lattice simulations of the baryon masses. Variations of the baryon masses along suitable pathes in the pion-kaon mass plane
are important to improve the determination of the low-energy parameters of QCD as encoded into the chiral Lagrangian.

Our work is based on the relativistic chiral Lagrangian with baryon octet and decuplet fields where effects at
N$^3$LO (next-to-next-to-next-leading order) are considered systematically. Here the leading order (LO) corresponds
to the chiral limit of the baryon masses and the NLO effects are linear in the quark masses. The details of the approach are published
in  \cite{Semke2005,Semke2007,Semke2012}. The chiral extrapolation of baryon masses with strangeness content is controversially
discussed in the literature \cite{Jenkins1991,Frink2006,Semke2007,PACS-CS2008,MartinCamalich:2010fp,PhysRevD.81.014503,Geng:2011wq,MartinCamalich:2010zz,Semke2012,WalkerLoud:2011ab}. The convergence properties of a strict chiral expansion for the baryon masses with three light flavors are very poor, if existing at all for physical strange quark masses. Different strategies how to perform partial summations or phenomenological adaptations are investigated by several groups \cite{Semke2005,Semke2007,Frink2006,PhysRevD.81.014503,MartinCamalich:2010fp}.
A straight forward application of chiral perturbation theory to recent QCD lattice simulations appears futile (see e.g.  \cite{LHPC2008,PACS-CS2008,MartinCamalich:2010fp,WalkerLoud:2011ab}).
A crucial element of our scheme is the self consistent determination of the baryon masses, i.e. the masses used in the one-loop contribution to the baryon self energies coincide with the physical masses defined by the pole position of the baryon propagator.
Furthermore, the low-energy constants required at N$^3$LO are estimated by sum rules that follow from QCD in the limit of a large number of
colors ($N_c$) \cite{LutzSemke2010,Semke2012}. Our approach was successfully tested against the available lattice data on the nucleon and
omega masses of the BMW group \cite{BMW2008}. Adjusting eight low-energy constants to the empirical baryon masses we fitted the remaining 6 parameters to the BMW data. In turn we obtained results that are in qualitative agreement with the predictions of the HSC group \cite{HSC2008}.
We challenge our approach further against recent lattice data from the PACS-CS and
LHPC groups \cite{PACS-CS2008,LHPC2008,Jenkins:2009wv,WalkerLoud:2011ab}, which assumed strange-quark masses significantly larger than the physical one.
Furthermore, we aim at a quantitative description of the HSC data taking into account their slightly too small strange quark mass.

Since we obtain a simultaneous and quantitative description of the lattice data of the BMW, PACS-CS, LHPC and HSC groups, we find it justified
to present detailed results on the strange-quark mass dependence of all members of the baryon octet and decuplet states. In particular, we confront
our parameter set against recent analyses of the BMW and QCDSF-UKQCD groups on the pion- and strangeness-sigma terms of the baryon octet
states \cite{Durr:2011mp,Horsley:2011wr,Bali:2011ks}.

\section{Chiral extrapolation of baryon masses}

We consider the chiral extrapolation of the baryon masses to unphysical quark masses. Assuming exact isospin symmetry, the hadron
masses are functions of  $m_u=m_d\equiv m_q$ and $m_s$. The dependence on the light quark masses may be traded against a dependence
on the pion and kaon masses. It is conventional to display the baryon masses as a function of the squared pion mass at fixed strange
quark mass. The 'physical' strange quark mass is determined such that at the physical pion mass the empirical kaon mass is reproduced.
Throughout this work we follow our approach as documented in  \cite{Semke2005,Semke2007,Semke2012}. In particular, we assume a quark-mass
dependence of the pion and kaon masses as predicted by $\chi$PT at the next-to-leading order with parameters as recalled in  \cite{Semke2012}.
The baryon masses are computed at N$^3$LO where we use physical baryon and meson masses in the one-loop contributions to the baryon self
energies and assume systematically large-$N_c$ sum rules for the parameter set.

\begin{table}[t]
\setlength{\tabcolsep}{4.5mm}
\renewcommand{\arraystretch}{1.3}
\begin{center}
\begin{tabular}{l||c|c|c|c|c}\hline
 & Fit 0 &  Fit 1 & Fit 2 & Fit 3 & Fit 4\\ \hline \hline
$\bar M_{[8]}\;\;$ [GeV] & \phantom{-}0.9111 & \phantom{-}0.8095 & \phantom{-}0.8004 & \phantom{-}0.8951 & \phantom{-}0.8745 \\
$\bar M_{[10]}$ [GeV]& \phantom{-}1.0938 &   \phantom{-}0.9508 & \phantom{-}1.1570 & \phantom{-}1.0961 & \phantom{-}1.1041 \\
$\bar b_0\,  \mathrm{[GeV^{-1}]}$ & -0.9086 & -0.9449 & -0.9722 & -0.8559 & -0.8792 \\
$\bar b_D\,  \mathrm{[GeV^{-1}]}$ & \phantom{-}0.5674 & \phantom{-}0.4451 & \phantom{-}0.4778 & \phantom{-}0.4623 & \phantom{-}0.4630 \\
$\bar b_F\,  \mathrm{[GeV^{-1}]}$ & -0.5880 & -0.4799 & -0.5197 & -0.4959 & -0.4952\\
$\bar d_0\,  \mathrm{[GeV^{-1}]}$ & -0.2300 &  -0.3902 & -0.1247 & -0.2082& -0.2070\\
$\bar d_D\,  \mathrm{[GeV^{-1}]}$ & -0.3617 & -0.4044 & -0.4258 & -0.4009 & -0.3965\\
$\bar c_0\,  \mathrm{[GeV^{-3}]}$ & \phantom{-}0.0176 & \phantom{-}0.0108 & \phantom{-}0.0101 & \phantom{-}0.0165 & \phantom{-}0.0157 \\
$\bar c_4\,  \mathrm{[GeV^{-3}]}$ & -0.1659 & \phantom{-}0.3031 & \phantom{-}0.1273& \phantom{-}0.3341 &\phantom{-}0.2942 \\
$\bar c_5\,  \mathrm{[GeV^{-3}]}$ & -0.3320 & -0.7264 & -0.5005 & -0.7647& -0.7348\\
$\bar c_6\,  \mathrm{[GeV^{-3}]}$ & -1.2366 & -1.4208 & -1.2841 & -1.5495 & -1.4903\\
$\bar e_4\,  \mathrm{[GeV^{-3}]}$ & -0.2520 & -0.2089 & -0.5068 & -0.3618 & -0.3459\\
$\bar \zeta_0\,  \mathrm{[GeV^{-2}]}$ & \phantom{-}1.1279 & \phantom{-}0.9520 & \phantom{-}0.8778 & \phantom{-}1.0921 &  \phantom{-}1.0449 \\
$\bar \zeta_D\,  \mathrm{[GeV^{-2}]}$ & \phantom{-}0.2848 & \phantom{-}0.3360 & \phantom{-}0.3432 & \phantom{-}0.2928 & \phantom{-}0.3035\\
$\bar \zeta_F\,  \mathrm{[GeV^{-2}]}$ & -0.2221 & -0.2973 & -0.3068 & -0.2338 & -0.2491\\
$\bar \xi_0\,  \mathrm{[GeV^{-2}]}$   & \phantom{-}1.1964 &  \phantom{-}1.0558 & \phantom{-}1.0425 & \phantom{-}1.1753 & \phantom{-}1.1477\\
\hline
\end{tabular}
\caption{The parameters are adjusted to reproduce the empirical values of the physical baryon octet and decuplet masses and various
lattice data as described in text.
}
\label{tab:parameters}
\end{center}
\end{table}

In our previous work \cite{Semke2012} we adjusted the parameter set to the experimental masses of the baryon octet
and decuplet states and to the results for the pion-mass dependence of the nucleon and omega masses as predicted
by the BMW group. Using the isospin average of the empirical baryon masses we derived a system of linear equations
which was used to express eight low-energy constants in terms of the remaining parameters. This provided a significant
simplification of the parameter determination and allowed us to analyze the set of non-linear and coupled equations in great depths.
An accurate reproduction of the physical baryon masses and lattice data was achieved.
All parameters, except of the symmetry preserving counter terms relevant at N$^3$LO, were considered.
Given the large-$N_c$ sum rules of  \cite{LutzSemke2010}, there are 5 independent parameters only, which were all set to zero.
The latter have a rather minor effect on the baryon masses and can be determined only with very precise lattice data.  For more details
of the fit strategy and the definition of the various parameters we refer to  \cite{Semke2012}. The parameters of that Fit 0 are recalled
in the 2nd column of Tab. 1. From the 16 parameters shown in Tab. 1 eight parameters  were used to recover the exact physical baryon masses. Two further parameter combinations were found to be redundant. That left six free parameters only to reproduce the eight lattice data of the BMW group.

While the results of the BMW group are for a physical strange quark mass, the results of the PACS-CS and LHPC groups rest on significantly
larger strange quark masses. The HSC group uses strange quark masses that are slightly smaller than the physical one. Given a
lattice data point with specified pion and kaon mass, we determine the corresponding quark masses, which are then used in the computation of
the baryon masses. In the following, we consider three scenarios leading to three distinct fit strategies.
Like in our previous works \cite{Lutz2002a,Semke2005,Semke2007,Semke2012}, we use  $F\simeq 0.45 $ and $D \simeq 0.80$
together with the values $H= 9\,F-3\,D $ and  $C=2\,D $ as implied by large-$N_c$ sum rules and $f= 92.4$ MeV
for the chiral limit value of the pion decay constant.

In our first scenario we supplement the BMW data with the recent results
from the LHPC group as presented in  \cite{WalkerLoud:2011ab}. As suggested in this work, it may be advantageous
to consider the following three particular mass combinations
\begin{eqnarray}
&& \!\!\!\!\!\!R_1 = \frac{5}{48}\, \Big(M_N + M_\Lambda + 3\,M_\Sigma + 2\,M_\Xi\Big) - \frac{1}{60}\,\Big(4\,M_\Delta + 3\,M_{\Sigma^*}
+ 2\,M_{\Xi^*} + M_\Omega\Big)\,, \nonumber \\
&&\!\!\!\!\!\! R_3 = \frac{5}{78}\,\Big(6\,M_N + M_\Lambda - 3\,M_\Sigma - 4\,M_\Xi\Big) - \frac{1}{39}\,\Big(2\,M_\Delta
- M_{\Xi^*} - M_\Omega\Big)\, , \nonumber \\
&& \!\!\!\!\!\! R_4 = \frac{1}{6} \,\Big(M_N + M_\Lambda - 3\,M_\Sigma + M_\Xi \Big)\,,
\label{def:mass_relations}
\end{eqnarray}
for which an improved analysis was provided. In Tab. \ref{tab:LHPC}  we recall the values for
the three mass relations in  (\ref{def:mass_relations}) at three different sets of pion and kaon masses from  \cite{WalkerLoud:2011ab}.
We included the nine lattice data points
into our $\chi^2$ function and obtained the parameter set shown in the third column of Tab. 1. Like for our previous results, the parameter set
is a consequence of adjusting six free parameters only to the considered lattice data. The physical baryon masses are reproduced exactly.
While the BMW data are still described with $\chi^2/N \simeq 0.7$, the LHPC data are described by our Fit 1 with $\chi^2/N \simeq 9.2$. We note that
the rather small uncertainties in the lattice predictions, as recalled in Tab. \ref{tab:LHPC} from  \cite{WalkerLoud:2011ab}, are based on a
single-lattice spacing  and do not yet consider effects from finite lattice volume corrections. The largest discrepancy of any of
our masses as compared to the corresponding lattice result is decreasing with increasing pion masses. For the smallest pion mass, where one would
expect the largest corrections from finite volume effects, the discrepancy is less than 18 MeV. If we assign an ad-hoc systematical
error of the form
\begin{eqnarray}
\Delta = 3\,\frac{e^{-L\,m_\pi}}{L\,m_\pi}\,{\rm GeV} \,,
\label{def-finite-volume}
\end{eqnarray}
we find for the LHPC data $\chi^2/N \simeq 2.9$ with $L\simeq 2.5$ fm of  \cite{LHPC2008}. The rough estimate (\ref{def-finite-volume}) of finite volume
effects is in line with the recent results of \cite{Procura:2006bj,Geng:2011wq,PhysRevD.84.014507}. We conclude that the BMW and LHPC data sets can both be described  by one
parameter set. We checked that, though already our original parameter set leads to a qualitative description of the LHPC data, a combined fit
improves the description of the LHPC data significantly without affecting the excellent reproduction of the BMW results. This reflects the fact
that the BMW data alone do not define a steep minimum in our $\chi^2$ function.

\begin{table}[t]
\begin{center}
\begin{tabular}{l||ll|ll|ll}\hline
$m_\pi$ & \multicolumn{2}{c|}{${3\over 2} R_1^{lat}$} & \multicolumn{2}{c|}{$R_3^{lat}$} & \multicolumn{2}{c}{$R_4^{lat}$} \\
$m_K$ & ${3\over 2} R_1^{\mathrm{\,Fit 3}}$ & $ {3\over 2} R_1^{\mathrm{\,Fit 1}}$ & $R_3^{\mathrm{\,Fit 3}}$ &
$R_3^{\mathrm{\,Fit 1}}$ & $R_4^{\mathrm{\,Fit 3}}$ & $R_4^{\mathrm{\,Fit 1}}$ \\ \hline \hline
320(2)& \multicolumn{2}{c|}{1285(6)}  & \multicolumn{2}{c|}{-113(3)} & \multicolumn{2}{c}{-39(2)} \\
640(2) & 1270 & 1277 & -130 & -131 & - 41 & -40 \\ \hline
389(2)& \multicolumn{2}{c|}{1315(6)}  & \multicolumn{2}{c|}{-100(2)} & \multicolumn{2}{c}{-33(1)} \\
659(2) & 1312 & 1317 & -109 & -109 & -33 & -32 \\ \hline
557(2)& \multicolumn{2}{c|}{1454(6)}  & \multicolumn{2}{c|}{-64(1)} & \multicolumn{2}{c}{-19(1)} \\
726(2) & 1461 & 1452 & -58 & -60 & -20 & -21 \\ \hline
\end{tabular}
\caption{Mass relations $R_1, R_3$ and $R_4$ in units of MeV for different values of pion and kaon masses. The lattice results
are taken from  \cite{WalkerLoud:2011ab}.}
\label{tab:LHPC}
\end{center}
\end{table}

\begin{table}[t]
\begin{center}
\begin{tabular}{l||ll|ll|ll|ll}\hline
$m_\pi$ & \multicolumn{2}{c|}{$M_N^{lat}$} & \multicolumn{2}{c|}{$M_\Lambda^{lat}$} & \multicolumn{2}{c|}{$M_\Sigma^{lat}$} & \multicolumn{2}{c}{$M_\Xi^{lat}$} \\
$m_K$ & $M_N^{\mathrm{Fit 3}}$ & $ M_N^{\mathrm{Fit 2}}$ & $M_\Lambda^{\mathrm{Fit 3}}$ & $M_\Lambda^{\mathrm{Fit 2}}$ & $M_\Sigma^{\mathrm{Fit 3}}$ & $M_\Sigma^{\mathrm{Fit 2}}$  & $M_\Xi^{\mathrm{Fit 3}}$ & $M_\Xi^{\mathrm{Fit 2}}$ \\ \hline \hline
155(6)& \multicolumn{2}{c|}{929(78)}  & \multicolumn{2}{c|}{1136(21)} & \multicolumn{2}{c|}{1215(21)} & \multicolumn{2}{c}{1389(7)} \\
552(2) & 949 & 955 & 1163 & 1164 & 1260 & 1264 & 1408 & 1402 \\ \hline
295(3)& \multicolumn{2}{c|}{1090(19)}  & \multicolumn{2}{c|}{1250(14)} & \multicolumn{2}{c|}{1311(15)} & \multicolumn{2}{c}{1443(10)} \\
592(2) & 1059 & 1073 & 1240 & 1245 & 1317 & 1320 & 1446 & 1439 \\ \hline
410(2) & \multicolumn{2}{c|}{1211(11)}  & \multicolumn{2}{c|}{1346(8)} & \multicolumn{2}{c|}{1396(8)} & \multicolumn{2}{c}{1498(7)} \\
633(1) & 1195 & 1215 & 1332 & 1341 & 1390 & 1392 & 1489 & 1484 \\ \hline \hline
383(3) & \multicolumn{2}{c|}{1156(15)}  & \multicolumn{2}{c|}{1270(9)} & \multicolumn{2}{c|}{1312(10)} & \multicolumn{2}{c}{1404(7)} \\
580(2) & 1156 & 1170 & 1274 & 1282 & 1323 & 1329 & 1412 & 1411 \\ \hline

\hline  \hline

 & \multicolumn{2}{c|}{$M_\Delta^{lat}$} & \multicolumn{2}{c|}{$M_{\Sigma^*}^{lat}$} & \multicolumn{2}{c|}{$M_{\Xi^*}^{lat}$} & \multicolumn{2}{c}{$M_\Omega^{lat}$} \\
 & $M_\Delta^{\mathrm{Fit 3}}$ & $ M_\Delta^{\mathrm{Fit 2}}$ & $M_{\Sigma^*}^{\mathrm{Fit 3}}$ & $M_{\Sigma^*}^{\mathrm{Fit 2}}$ & $M_{\Xi^*}^{\mathrm{Fit 3}}$ & $M_{\Xi^*}^{\mathrm{Fit 2}}$  & $M_\Omega^{\mathrm{Fit 3}}$ & $M_\Omega^{\mathrm{Fit 2}}$ \\ \hline \hline
155(6)& \multicolumn{2}{c|}{1257(82)}  & \multicolumn{2}{c|}{1493(30)} & \multicolumn{2}{c|}{1637(15)} & \multicolumn{2}{c}{1766(7)} \\
552(2) & 1245 & 1241 & 1433 & 1428 & 1616 & 1606 & 1782 & 1776 \\ \hline
295(3)& \multicolumn{2}{c|}{1396(20)}  & \multicolumn{2}{c|}{1539(15)} & \multicolumn{2}{c|}{1678(13)} & \multicolumn{2}{c}{1809(11)} \\
592(2) & 1362 & 1364 & 1524 & 1520 & 1678 & 1668 & 1819 & 1808 \\ \hline
410(2) & \multicolumn{2}{c|}{1508(14)}  & \multicolumn{2}{c|}{1619(9)} & \multicolumn{2}{c|}{1727(9)} & \multicolumn{2}{c}{1834(8)} \\
633(1) & 1497 & 1501 & 1615 & 1616 & 1737 & 1728 & 1846 & 1838 \\ \hline \hline
383(3) & \multicolumn{2}{c|}{1460(19)}  & \multicolumn{2}{c|}{1550(16)} & \multicolumn{2}{c|}{1643(13)} & \multicolumn{2}{c}{1735(11)} \\
580(2) & 1463 & 1461 & 1565 & 1561 & 1667 & 1658 & 1762 & 1752 \\ \hline
\end{tabular}
 \caption{Baryon masses in units of MeV. The lattice results are taken from  \cite{PACS-CS2008}. }
\label{tab:PACS}
\end{center}
\end{table}

Our second scenario considers the combined results of the BMW and PACS-CS groups. In Tab. \ref{tab:PACS}
we recall the results of the PACS-CS group at 4 different sets of pion and kaon masses. We added the contribution of the 32 lattice
points to the $\chi^2$ function and searched for its minimum. The resulting parameters are collected under Fit 2 in Tab. 1.
This fit  describes the PACS-CS data with $\chi^2/N \simeq 1.4$. The quality of the description of the BMW data is still excellent with a
$\chi^2/N \simeq 0.8$. It is interesting to compare the size of the various parameters of the three fits 0, 1 and 2. All parameters
have quite similar values suggesting a high level of compatibility of the different lattice data sets.

In our third scenario we consider the combined lattice results of the BMW, LHPC and PACS-CS groups. The minimum of our $\chi^2$ function
is obtained with the parameter set of Fit 3 in Tab. 1. Again, the empirical masses are reproduced exactly. The minimum of the
$\chi^2$ function is reached by the variation of six free parameters only. This fit describes the LHPC data with $\chi^2/N \simeq 3.9$, the PACS-CS data with $\chi^2/N \simeq 1.7$ and the BMW data with $\chi^2/N \simeq 0.7$. Detailed results are included in Tab.  \ref{tab:LHPC} and \ref{tab:PACS}.
We find this an encouraging result giving us high confidence in the extracted low-energy constants as
collected in Tab. 1.

\begin{figure}
\centering
\includegraphics[width=11.0cm,clip=true]{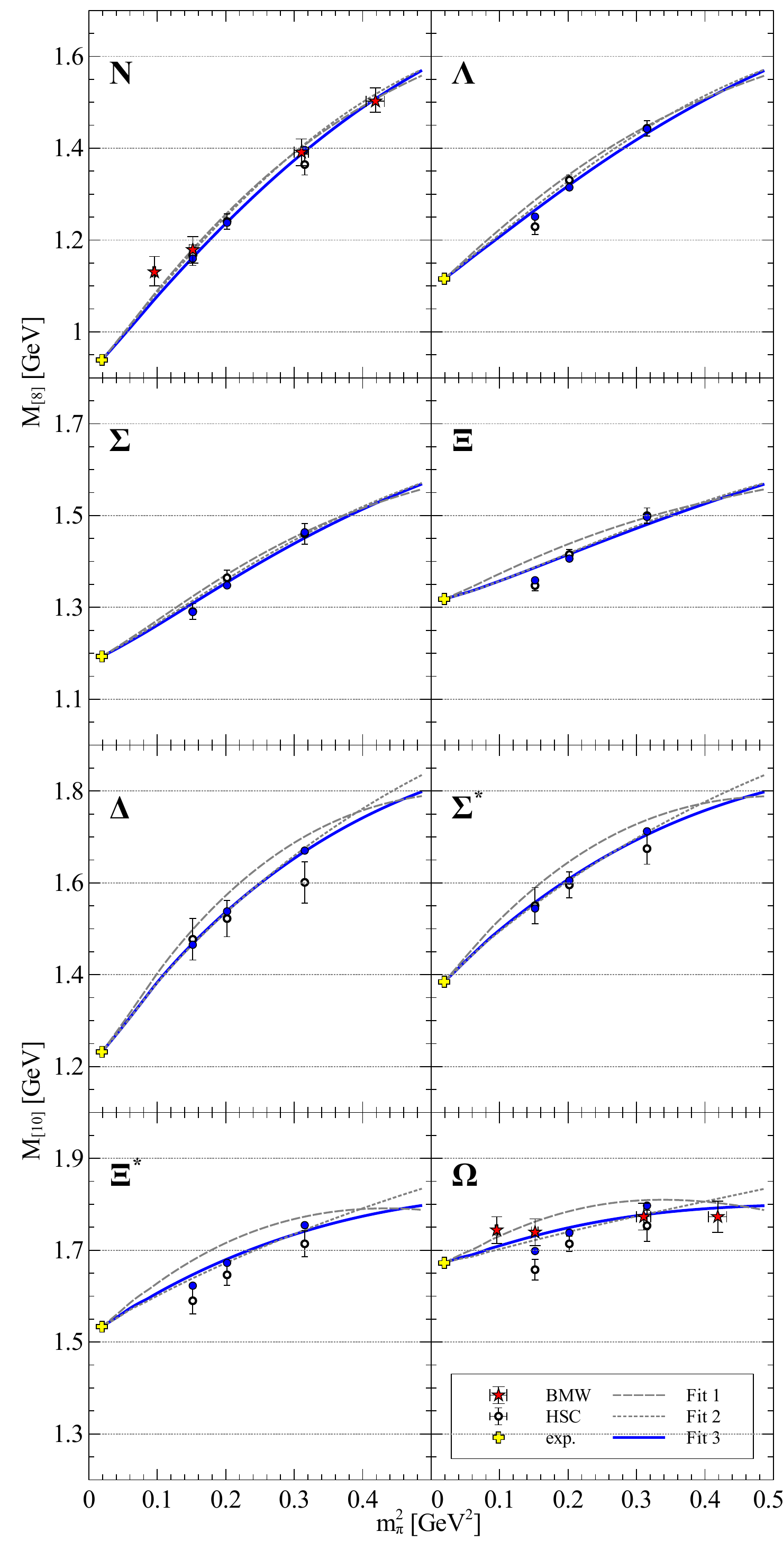}
\caption{Baryon masses as a function of the squared pion mass at the physical strange quark mass as explained in the text.
The HSC data were not included in Fit 1-3.}
\label{figure:baryon-masses}
\end{figure}

In Fig. \ref{figure:baryon-masses} we show the pion-mass dependence of the baryon masses as implied by the first three
parameter sets of Tab. 1. The solid lines follow with parameter set 3,  the dotted lines with set 2 and the dashed lines with set 1.
The results are confronted against the lattice data from the BMW and HSC groups. The spread in the
three lines is comfortably small. We recall that the HSC data rely on a slightly too small strange quark mass. Therefore,
the HSC data points should not be compared quantitatively with any of the lines in Fig. 1. Our three line types assume a physical strange
quark mass as required by a comparison with the BMW data points. In order to provide a quantitative comparison, we
compute the baryon masses for the pion and kaon masses as assumed by the HSC group. The filled circles show our results based on
the parameter set 3. The distance of the filled circles form the solid lines in Fig.  \ref{figure:baryon-masses} measures the
importance of taking the precise physical strange quark mass in the computation of the baryon masses. Most prominent is the effect in
the pion-mass dependence of the omega mass, where the filled circles are reasonably close to the predictions of the HSC group. The HSC data set
is described with a  $\chi^2/N \simeq 1.0$. Assuming parameter set 2, we obtain an
almost as good description with $\chi^2/N \simeq 1.1$. The parameter set 1 leads to a significantly worse description with
$\chi^2/N \simeq 3.5$.

\begin{figure}
\centering
\includegraphics[width=8.5cm,clip=true]{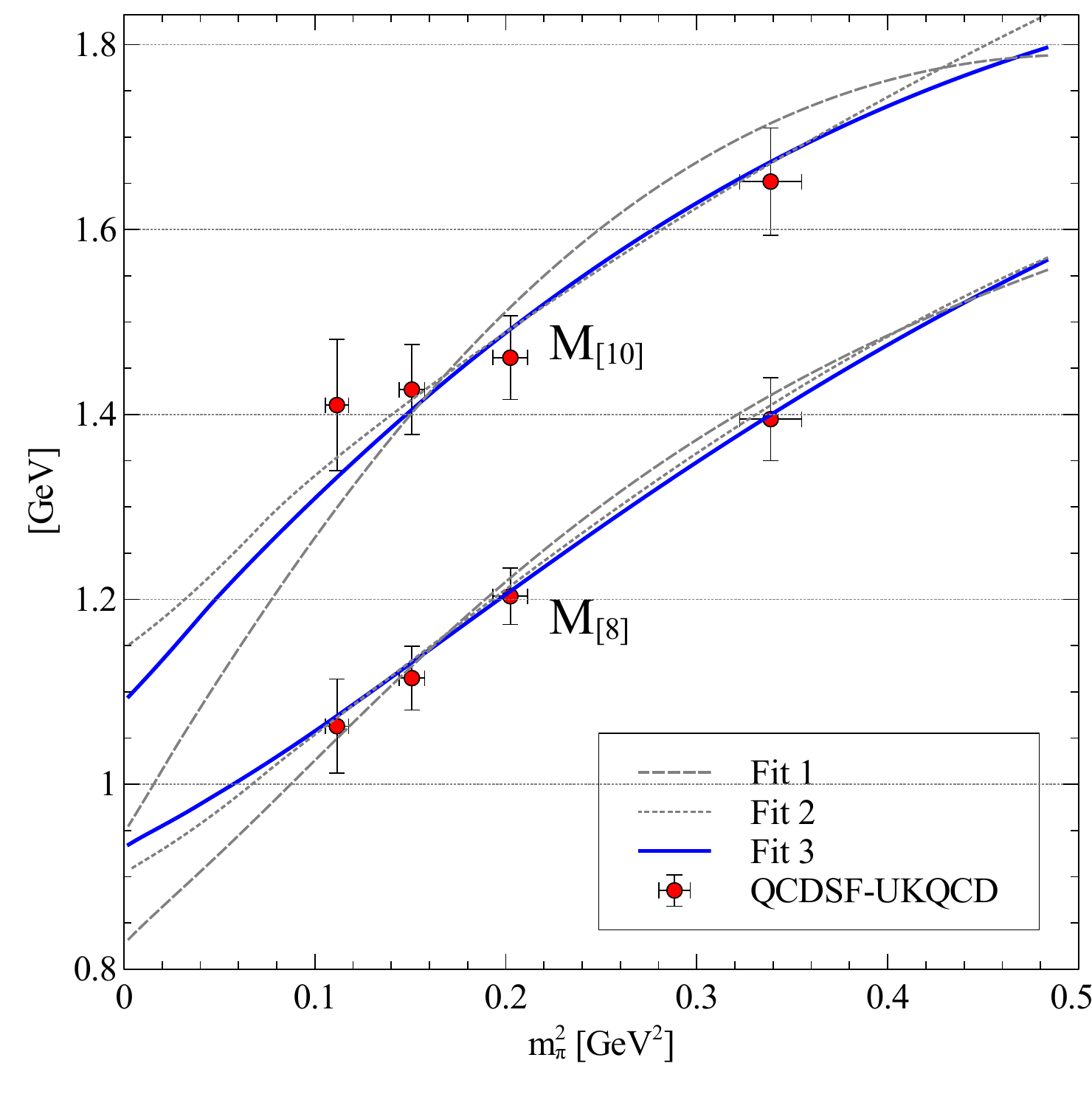}
\caption{Pion-mass extrapolation of the baryon octet and decuplet masses in the flavor symmetric limit.
The QCDSF-UKQCD data were not included in Fit 1-3.}
\label{figure:flavor_symmetric_case}
\end{figure}

\begin{figure}
\centering
\includegraphics[width=11.0cm,clip=true]{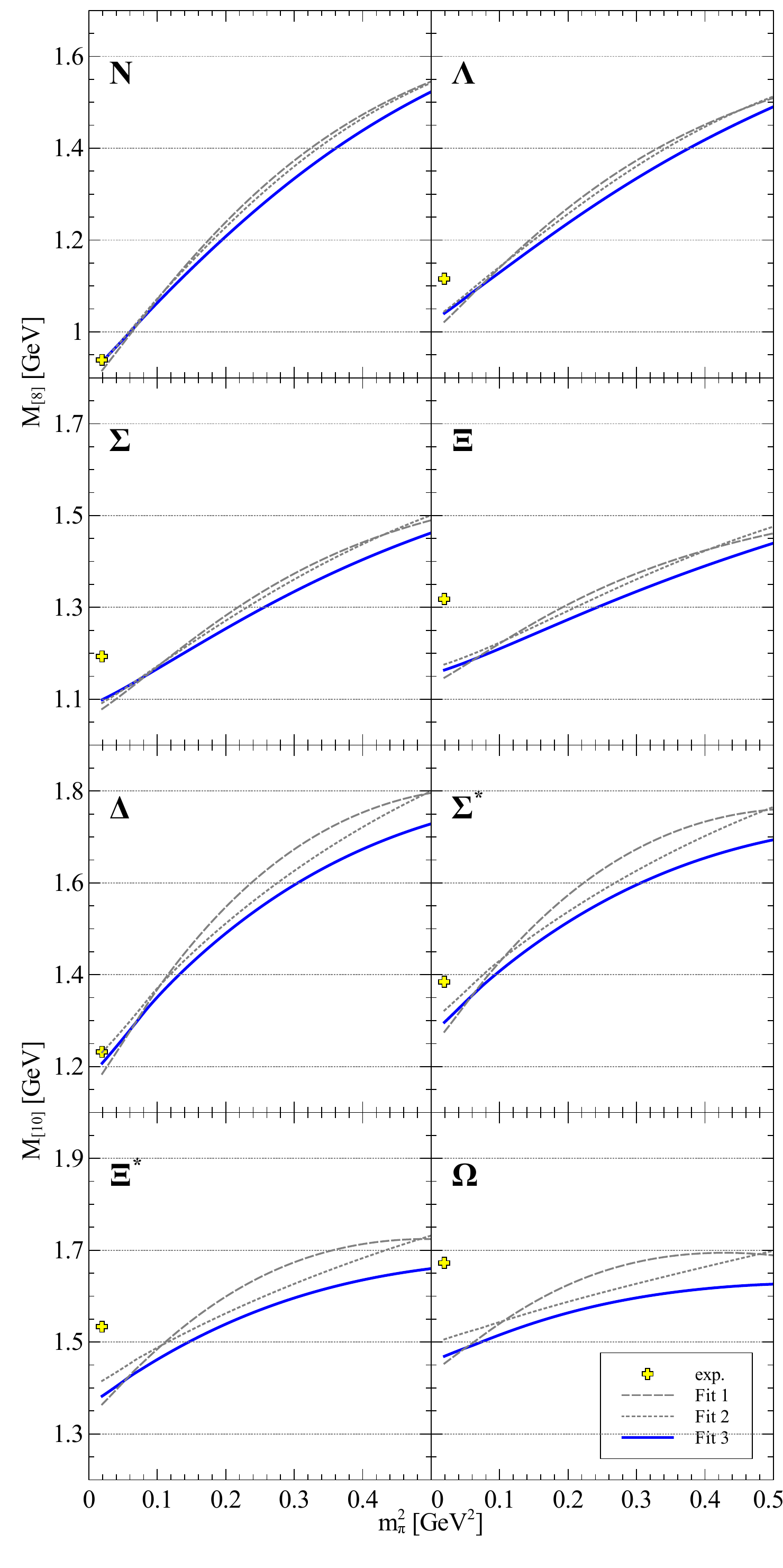}
\caption{Baryon masses as a function of the squared pion mass at an unphysical strange quark mass as explained in the text.  }
\label{figure:baryon-masses-predictions}
\end{figure}

In the remaining part of this section we wish to explore how to further improve the determination of the low-energy constants. This
is a crucial  issue in view of the additional five symmetry conserving parameters that were not considered so far. We feel that an
inclusion of those five parameters in our $\chi^2$ function would not be significant at this stage. The total $\chi^2/N \simeq 1.65 $ computed
for all lattice points considered here is too close to one. A more detailed estimate of systematic
lattice effects like finite volume effects or the influence of smaller lattice spacings would be necessary.
Given the small spread in our predictions for the pion-mass dependence of the baryon masses at physical strange quark mass, we find
it useful to consider the flavor symmetric extrapolation with $m_q = m_s$, where the baryons form a degenerate octet and decuplet.
In Fig. \ref{figure:flavor_symmetric_case} the octet and decuplet masses are shown as a function of the pion mass. In the limit of
vanishing pion masses the figure predicts the three-flavor chiral limit value of the baryon  masses. These values
are of crucial importance in any effective field theory approach to hadron physics involving strangeness.
The Fig. \ref{figure:flavor_symmetric_case}  includes recent results of the QCDSF-UKQCD Collaboration in the flavor symmetric limit \cite{Bietenholz:2011qq}. Given the fact that these lattice data were not fitted, the agreement with the results of Fit 2 or 3 is remarkable.
We suggest that QCD lattice collaborations perform further simulations for such a flavor symmetric case.

We continue the discussion of the chiral extrapolation by showing in Fig. \ref{figure:baryon-masses-predictions} our predictions for the pion-mass dependence of the baryon masses at an unphysical strange quark mass. At the physical pion mass our particular choice implies a kaon mass of 400 MeV. Our results suggest that such lattice simulations, in particular of any the decuplet masses, would be quite discriminative and therefore extremely useful in the path of establishing QCD's low energy parameters.

Finally, we investigated the role of the parameters $F$ and $D$. While the first three fits were obtained with $F=0.45$ and $D=0.80$,
we allowed for a variation of those parameters in an additional Fit 4. A combined fit to the lattice data of the BMW, LHPC, PACS-CS and HSC groups was performed leading to the parameters collected in Tab. 1 together with $F=0.452$ and $D=0.799$. This fit describes the LHPC data with $\chi^2/N \simeq 4.3$, the PACS-CS data with $\chi^2/N \simeq 1.7$, the BMW data with $\chi^2/N \simeq 0.7$ and the HSC data with $\chi^2/N \simeq 0.8$. A similar fit excluding the HSC data leads to $F = 0.444$ and $D = 0.714$, however, the implied $\chi^2/N \simeq 6.5$ for the HSC data is significantly worse.
Comparable values for $F$ and $D$ were obtained previously in \cite{WalkerLoud:2011ab} based on an analysis of the LHPC data only.
We conclude that using the complete set of lattice data the coupling constants $F$ and $D$ can be determined accurately. We refrain from including the results of this Fit 4 into the Figs. 1-3, since the corresponding lines are almost indistinguishable from the results of our Fit 3.

\section{Pion and strangeness baryon sigma terms}

The pion- and strangeness baryon sigma terms play an important role in various physical systems.
For instance, the pion-nucleon sigma term is of greatest relevance in the determination of the
density dependence of the quark condensate at low baryon densities and therefore
provides a first estimate of the critical baryon density at which chiral symmetry may be restored
(see e.g.  \cite{Lutz:1999vc}).  Similarly, the kaon-nucleon sigma terms are key parameters for the
determination of a possible kaon condensate in dense nuclear matter \cite{Kaplan:1986yq}.

Assuming exact isospin symmetry with $m_u=m_d\equiv m_q$,  the pion-nucleon sigma term reads
\begin{eqnarray}
\sigma_{\pi N} = m_q\,\langle N(p)\,| \bar q\, q | N(p) \rangle = m_q\,\frac{\partial}{\partial m_q} m_N\,.
\label{def-sigmapiN}
\end{eqnarray}
The pion-sigma terms of the remaining baryon states are defined analogously to  (\ref{def-sigmapiN}).
As indicated in  (\ref{def-sigmapiN}), the matrix elements of the scalar quark operator are accessible via the derivative of the nucleon mass with respect to the light quark mass $m_q$. This follows directly from the Feynman-Hellman theorem.
The strangeness sigma term $\sigma_{sN}$ of the nucleon is determined by the derivative of
the nucleon mass with respect to the strange quark mass $m_s$:
\begin{eqnarray}
\sigma_{s N} = m_s\,\langle N(p)\,| \bar s\, s | N(p) \rangle = m_s\,\frac{\partial}{\partial m_s} m_N\,.
\label{def-sigmasN}
\end{eqnarray}

\begin{table}[b]
\begin{center}
\begin{tabular}{l||l|c|c|cccc}\hline
 &  \cite{Durr:2011mp} &  \cite{Horsley:2011wr} &  \cite{MartinCamalich:2010fp} & Fit 1 & Fit 2 & Fit 3 & Fit 4\\ \hline \hline
$\sigma_{\pi N}$ & $39(4) ^{+ 18}_{- 7}  $ &$ 31 (3)(4)$ & 59(2)(17)  & 34 & 33 & 31  &31\\
$\sigma_{\pi \Lambda}$ & $29(3)  ^{+11}_{-5} $ &24(3)(4) & 39(1)(10) & 24 & 21 & 20  & 20\\
$\sigma_{\pi \Sigma}$ & $ 28(3) ^{+19}_{-3}$ & 21(3)(3)& 26(2)(5) &17 & 15 & 14 & 14\\
$\sigma_{\pi  \Xi}$ & $16(2)^{+8}_{-3}  $&16(3)(4) & 13(2)(1) & 11 & 7 & 7  & 7\\ \hline \hline
$\sigma_{s N}$ & $\;\;67(27)^{+55}_{-47}  $ &$ 71(34)(59)$ & -4(23)(25) & 43 & 24 & 2 & 3\\
$\sigma_{s \Lambda}$ & $180(26)  ^{+48}_{-77} $ &247(34)(69)  & 126(26)(35) & 238 & 194 & 191 & 194 \\
$\sigma_{s \Sigma}$ & $ 245(29) ^{+50}_{-72}$ & 336(34)(69)& 159(27)(45) & 311 & 291 & 273 & 278\\
$\sigma_{s  \Xi}$ & $312(32)^{+72}_{-77}  $&468(35)(59) & 267(31)(50) & 449  & 380 & 407 & 408 \\\hline
\end{tabular}
\caption{Pion- and strangeness sigma terms of the baryon octet states in units of MeV. }
\label{tab:sigmaterms_octet}
\end{center}
\end{table}

In Tab. \ref{tab:sigmaterms_octet} we present our predictions for the pion- and strangeness sigma terms of the baryon octet states.
They are compared with two recent lattice determinations \cite{Durr:2011mp,Horsley:2011wr}. Our values for the sigma terms are
in reasonable agreement with the lattice results. We find it encouraging that our values are compatible with both lattice groups
within one sigma deviation in almost all cases. In particular, we obtain a rather small value for the pion-nucleon sigma term, which is
within reach of the seminal result $\sigma_{\pi N}= 45 \pm 8$ MeV of Gasser, Leutwyler and Sainio in  \cite{Gasser:1990ce}.
The size of the pion-nucleon term can be determined from the pion-nucleon scattering data. It requires a subtle subthreshold
extrapolation of the scattering data. Despite the long history of the sigma-term physics, the precise determination is still
highly controversial (for one of the first reviews see e.g.  \cite{Reya:1974gk}). Such a result is also consistent with
the recent analysis of the QCDSF collaboration \cite{Bali:2011ks}, which suggests a value $\sigma_{\pi N} =38 \pm 12$ MeV.
Our estimate for the strangeness sigma term of the nucleon with $\sigma_{sN} \simeq 22 \pm 20 $ MeV
is compatible with the currently most precise lattice prediction $\sigma_{sN}= 12^{+23}_{-16}$ MeV in  \cite{Bali:2011ks}.

\begin{table}[t]
\begin{center}
\begin{tabular}{l|l|cccc}\hline
 &  \cite{MartinCamalich:2010fp}  & Fit 1 & Fit 2 & Fit 3  & Fit 4\\ \hline \hline
$\sigma_{\pi \Delta}$ & $55(4)(18)  $ & 37 & 32 & 33 & 32\\
$\sigma_{\pi \Sigma^*}$ & $39(3)(13) $ & 31 & 26 & 26 & 26\\
$\sigma_{\pi \Xi^*}$ & $22(3)(7)$ & 22 & 15 & 17 & 16\\
$\sigma_{\pi  \Omega}$ & $\;\;5(2)(1)  $ & 14 & 7 & 9 & 9 \\
 \hline \hline
$\sigma_{s \Delta}$ & $\;\;56(24)(1) $ & 82 & 0 & 28 & 16\\
$\sigma_{s \Sigma^*}$ & $160(28)(7) $ & 255 & 167 & 200 & 191\\
$\sigma_{s \Xi^*}$ & $ 274(32)(9)$ & 425 & 320 & 371 & 363\\
$\sigma_{s  \Omega}$ & $360(34)(26)  $ & 559 & 460 & 503 & 496 \\ \hline
\end{tabular}
\caption{Pion and strangeness sigma terms of the baryon decuplet states in units of MeV.}
\label{tab:sigmaterms_decuplet}
\end{center}
\end{table}

In Tab. \ref{tab:sigmaterms_octet}  we recall also the results of a chiral extrapolation study of the recent PACS-CS data by Camalich et al.
\cite{MartinCamalich:2010fp}. Similar sigma terms were obtained previously  by Young and Thomas \cite{Young2009}. Both analyses  are based on the baryon masses truncated at N$^2$LO. While Camalich et al. work with phenomenologically adjusted values for the meson-baryon coupling constants, the
scheme of Young and Thomas relies on various phenomenological form factors in the meson-baryon loop integrals.
For almost all sigma terms we find significant differences to their results. This may reflect the much less accurate reproduction of the PACS-CS data and the empirical baryon masses in  \cite{Young2009,MartinCamalich:2010fp}.

We turn to the sigma terms of the decuplet states, for which our predictions are compiled in Tab. \ref{tab:sigmaterms_decuplet}.
Again, we compare our values with the ones obtained in  \cite{MartinCamalich:2010fp}. Like in the case of the baryon octet states there are
significant differences.

\section{Summary and outlook}

In this work we presented the first simultaneous and quantitative description of the baryon octet and decuplet masses as computed by
four different lattice groups, BMW, LHPC, PACS-CS and HSC. Using the chiral Lagrangian at N$^3$LO we obtained a universal parameter set that
leads to a quantitative reproduction of the baryon masses and the lattice simulation data at different pairs of pion and kaon masses. While
the experimental masses are reproduced exactly, the lattice data are successfully fitted with six free parameters only. The total $\chi^2/N \simeq 1.65$
of all considered lattice data is amazingly small. As a further consistency check we computed various flavor symmetric limits of the baryon masses
as obtained recently by the QCDSF-UKQCD group. Their values are consistent with our results.
This suggests a high level of compatibility of the different lattice data sets.
In addition we explored the role of the axial coupling constants $F$ and $D$, which we found to be accurately determined from
a simultaneous fit of the lattice data.

Based on this result, we predicted the pion and strangeness sigma terms of all baryon ground states.
Assuming different lattice data as input, we find rather small variations in our predictions only. In particular, we obtain a quite small pion-nucleon sigma term of about $32 \pm 2$ MeV and  also for the strangeness sigma term of the nucleon of about $22\pm 20$ MeV .

In order to further diminish the residual uncertainties in the low-energy constants of QCD, we presented predictions for the pion-mass
dependence of the baryon masses in the  flavor symmetric limit and for a somewhat reduced strange quark mass. Future lattice simulations
along those pathes in the pion-kaon plane would help to consolidate our parameter set and may lead to the determination of the
additional five flavor symmetric counter terms that were not considered yet. The latter play a decisive role in meson-baryon scattering
processes at next-to-leading order already.

\bibliography{literatur}{}

\begin{thebibliography}{10}
\expandafter\ifx\csname url\endcsname\relax
  \def\url#1{\texttt{#1}}\fi
\expandafter\ifx\csname urlprefix\endcsname\relax\def\urlprefix{URL }\fi

\bibitem{MILC2004}
C.~Aubin, et~al., {Light hadrons with improved staggered quarks: Approaching
  the continuum limit}, Phys. Rev. D70 (2004) 094505.

\bibitem{LHPC2008}
A.~Walker-Loud, et~al., {Light hadron spectroscopy using domain wall valence
  quarks on an Asqtad sea}, Phys. Rev. D79 (2009) 054502.

\bibitem{PACS-CS2008}
S.~Aoki, et~al., {2+1 Flavor Lattice QCD toward the Physical Point}, Phys. Rev.
  D79 (2009) 034503.

\bibitem{HSC2008}
H.-W. Lin, et~al., {First results from 2+1 dynamical quark flavors on an
  anisotropic lattice: light-hadron spectroscopy and setting the strange-quark
  mass}, Phys. Rev. D79 (2009) 034502.

\bibitem{BMW2008}
S.~Durr, et~al., {Ab-Initio Determination of Light Hadron Masses}, Science 322
  (2008) 1224--1227.

\bibitem{Alexandrou:2009qu}
C.~Alexandrou, et~al., {Low-lying baryon spectrum with two dynamical twisted
  mass fermions}, Phys.Rev. D80 (2009) 114503.

\bibitem{Durr:2011mp}
S.~Durr, et~al., {Sigma term and strangeness content of octet baryons}, Phys.
  Rev. D85 (2012) 014509.

\bibitem{WalkerLoud:2011ab}
A.~Walker-Loud, {Evidence for non-analytic light quark mass dependence in the
  baryon spectrum}, arXiv:1112.2658 [hep-lat].

\bibitem{Semke2005}
A.~Semke, M.~F.~M. Lutz, {Baryon self energies in the chiral loop expansion},
  Nucl. Phys. A778 (2006) 153--180.

\bibitem{Semke2007}
A.~Semke, M.~F.~M. Lutz, {On the possibility of a discontinuous quark-mass
  dependence of baryon octet and decuplet masses}, Nucl. Phys. A789 (2007)
  251--259.

\bibitem{Semke2012}
A.~Semke, M.~F.~M. Lutz, {Quark-mass dependence of the baryon ground-state
  masses}, Phys. Rev. D85 (2012) 034001--034012.

\bibitem{Jenkins1991}
E.~Jenkins, A.~V. Manohar, Baryon chiral perturbation theory using a heavy
  fermion lagrangian, Phys. Lett. B 255 (1991) 558.

\bibitem{Frink2006}
M.~Frink, U.-G. Meissner, {On the chiral effective meson-baryon Lagrangian at
  third order}, Eur. Phys. J. A29 (2006) 255--260.

\bibitem{MartinCamalich:2010fp}
J.~Martin~Camalich, L.~S. Geng, M.~J. Vicente~Vacas, {The lowest-lying baryon
  masses in covariant SU(3)-flavor chiral perturbation theory}, Phys. Rev. D82
  (2010) 074504.

\bibitem{PhysRevD.81.014503}
R.~D. Young, A.~W. Thomas, Octet baryon masses and sigma terms from an su(3)
  chiral extrapolation, Phys. Rev. D 81 (2010) 014503.

\bibitem{Geng:2011wq}
L.-s. Geng, X.-l. Ren, J.~Martin-Camalich, W.~Weise, {Finite-volume effects on
  octet-baryon masses in covariant baryon chiral perturbation theory}, Phys.
  Rev. D84 (2011) 074024.

\bibitem{MartinCamalich:2010zz}
J.~Martin~Camalich, L.~Geng, M.~Vicente~Vacas, {Analysis of the LQCD results on
  the lowest-lying baryon masses in chiral perturbation theory}, AIP Conf.Proc.
  1322 (2010) 440--444.

\bibitem{LutzSemke2010}
M.~F.~M. Lutz, A.~Semke, {Large-Nc operator analysis of 2-body meson-baryon
  counterterms in the chiral Lagrangian}, Phys. Rev. D83 (2011) 034008.

\bibitem{Jenkins:2009wv}
E.~E. Jenkins, A.~V. Manohar, J.~W. Negele, A.~Walker-Loud, {A Lattice Test of
  1/N(c) Baryon Mass Relations}, Phys.Rev. D81 (2010) 014502.

\bibitem{Horsley:2011wr}
R.~Horsley, et~al., {Hyperon sigma terms for 2+1 quark flavours},
  arXiv:1110.4971 [hep-lat].

\bibitem{Bali:2011ks}
G.~S. Bali, et~al., {The strange and light quark contributions to the nucleon
  mass from Lattice QCD}, arXiv:1111.1600 [hep-lat].

\bibitem{Lutz2002a}
M.~F.~M. Lutz, E.~E. Kolomeitsev, {Relativistic chiral SU(3) symmetry, large
  $N_c$ sum rules and meson-baryon scattering}, Nucl. Phys. A700 (2002)
  193--308.

\bibitem{Procura:2006bj}
M.~Procura, B.~U. Musch, T.~Wollenweber, T.~R. Hemmert, W.~Weise, {Nucleon
  mass: From lattice QCD to the chiral limit}, Phys. Rev. D73 (2006) 114510.

\bibitem{PhysRevD.84.014507}
S.~R. Beane, et~al., High statistics analysis using anisotropic clover
  lattices: Iv. volume dependence of light hadron masses, Phys. Rev. D 84
  (2011) 014507.

\bibitem{Bietenholz:2011qq}
W.~Bietenholz, et~al., {Flavour blindness and patterns of flavour symmetry
  breaking in lattice simulations of up, down and strange quarks}, Phys.Rev.
  D84 (2011) 054509.

\bibitem{Lutz:1999vc}
M.~F.~M. Lutz, B.~Friman, C.~Appel, {Saturation from nuclear pion dynamics},
  Phys. Lett. B474 (2000) 7--14.

\bibitem{Kaplan:1986yq}
D.~Kaplan, A.~Nelson, {Strange Goings on in Dense Nucleonic Matter}, Phys.Lett.
  B175 (1986) 57--63.

\bibitem{Gasser:1990ce}
J.~Gasser, H.~Leutwyler, M.~Sainio, {Sigma term update}, Phys.Lett. B253 (1991)
  252--259.

\bibitem{Reya:1974gk}
E.~Reya, {Chiral symmetry breaking and meson - nucleon sigma commutators: A
  Review}, Rev. Mod. Phys. 46 (1974) 545--580.

\bibitem{Young2009}
R.~D. Young, A.~W. Thomas, {Octet baryon masses and sigma terms from an SU(3)
  chiral extrapolation}.

\end{thebibliography}
\bibliographystyle{elsart-num}

\end{document}